\documentstyle[12pt]{article}

\makeatletter
\def\eqnarray{\stepcounter{equation}\let\@currentlabel=\theequation
\global\@eqnswtrue
\global\@eqcnt\z@\tabskip\@centering\let\\=\@eqncr
$$\halign to \displaywidth\bgroup\@eqnsel\hskip\@centering
  $\displaystyle\tabskip\z@{##}$&\global\@eqcnt\@ne 
  \hfil$\displaystyle{\hbox{}##\hbox{}}$\hfil
  &\global\@eqcnt\tw@ $\displaystyle\tabskip\z@
  {##}$\hfil\tabskip\@centering&\llap{##}\tabskip\z@\cr}
\def\@sect#1#2#3#4#5#6[#7]#8{\ifnum #2>\c@secnumdepth
    \def\@svsec{}\else
    \refstepcounter{#1}\edef\@svsec{\csname the#1\endcsname.\hskip 1em }\fi
    \@tempskipa #5\relax
    \ifdim \@tempskipa>\z@
    \begingroup #6\relax
    \@hangfrom{\hskip #3\relax\@svsec}{\interlinepenalty \@M #8\par}
    \endgroup
    \csname #1mark\endcsname{#7}\addcontentsline
    {toc}{#1}{\ifnum #2>\c@secnumdepth \else
     \protect\numberline{\csname the#1\endcsname}\fi
           #7}\else
    \def\@svsechd{#6\hskip #3\@svsec #8\csname #1mark\endcsname
          {#7}\addcontentsline
          {toc}{#1}{\ifnum #2>\c@secnumdepth \else
     \protect\numberline{\csname the#1\endcsname}\fi
           #7}}\fi
     \@xsect{#5}}
\def\label#1{\@bsphack\if@filesw {\let\thepage\relax
   \xdef\@gtempa{\write\@auxout{\string
   \newlabel{#1}{{\thesection.\@currentlabel}{\thepage}}}}}\@gtempa
   \if@nobreak \ifvmode\nobreak\fi\fi\fi\@esphack}
\def\@eqnnum{(\thesection.\theequation)}
\def\section{\setcounter{equation}{0} \@startsection {section}{1}{\z@}{-3.5ex
   plus -1ex minus -.2ex}{2.3ex plus .2ex}{\Large\bf}}
\newcount\@minsofar
\newcount\@min
\newcount\@cite@temp
\def\@citex[#1]#2{%
\if@filesw \immediate \write \@auxout {\string \citation {#2}}\fi
\@tempcntb\m@ne \let\@h@ld\relax \def\@citea{}%
\@min\m@ne%
\@cite{%
  \@for \@citeb:=#2\do {\@ifundefined {b@\@citeb}%
    {\@h@ld\@citea\@tempcntb\m@ne{\bf ?}%
    \@warning {Citation `\@citeb ' on page \thepage \space undefined}}%
{\@minsofar\z@ \@for \@scan@cites:=#2\do {%
  \@ifundefined{b@\@scan@cites}%
    {\@cite@temp\m@ne}
    {\@cite@temp\number\csname b@\@scan@cites \endcsname \relax}%
\ifnum\@cite@temp > \@min
    \ifnum\@minsofar = \z@
      \@minsofar\number\@cite@temp
      \edef\@scan@copy{\@scan@cites}\else
    \ifnum\@cite@temp < \@minsofar
      \@minsofar\number\@cite@temp
      \edef\@scan@copy{\@scan@cites}\fi\fi\fi}\@tempcnta\@min
  \ifnum\@minsofar > \z@ 
    \advance\@tempcnta\@ne
    \@min\@minsofar
    \ifnum\@tempcnta=\@minsofar 
      \ifx\@h@ld\relax
        \edef \@h@ld{\@citea\csname b@\@scan@copy\endcsname}%
    \else \edef\@h@ld{\ifmmode{-}\else--\fi\csname b@\@scan@copy\endcsname}%
      \fi
    \else \@h@ld\@citea\csname b@\@scan@copy\endcsname
          \let\@h@ld\relax
  \fi 
\fi}%
\def\@citea{,\penalty\@highpenalty\,}}\@h@ld}{#1}}
\def\appendixname{Appendix}
\def\appendix{\par
  \def\pre@section{\appendixname{}}
  \setcounter{section}{1}
  \@addtoreset{equation}{section}
  \def\thesection{\Alph{section}}
  \def\theequation{\arabic{equation}}}

\def\appendix{\par
  \def\pre@section{\appendixname{}}
  \setcounter{section}{1}
  \@addtoreset{equation}{section}
  \def\thesection{\Alph{section}}
  \def\theequation{\arabic{equation}}}

\makeatother

\begin{document}
\addtolength{\unitlength}{-0.5\unitlength}
\newsavebox{\near}\savebox{\near}(30,30){\begin{picture}(30,30)
\thicklines\put(10,20){\line(1,-1){10}}\put(10,20){\line(2,1){20}}
\put(20,10){\line(1,2){10}}\end{picture}}

\newsavebox{\nwar}\savebox{\nwar}(30,30){\begin{picture}(30,30)
\thicklines\put(10,10){\line(1,1){10}}\put(10,10){\line(-1,2){10}}
\put(20,20){\line(-2,1){20}}\end{picture}}

\newsavebox{\swar}\savebox{\swar}(30,30){\begin{picture}(30,30)
\thicklines\put(10,20){\line(1,-1){10}}\put(0,0){\line(2,1){20}}
\put(0,0){\line(1,2){10}}\end{picture}}

\newsavebox{\sear}\savebox{\sear}(30,30){\begin{picture}(30,30)
\thicklines\put(10,10){\line(1,1){10}}\put(10,10){\line(2,-1){20}}
\put(20,20){\line(1,-2){10}}\end{picture}}

\newsavebox{\nar}\savebox{\nar}(30,30){\begin{picture}(30,30)
\thicklines\put(5,10){\line(1,0){20}}\put(5,10){\line(1,2){10}}
\put(15,30){\line(1,-2){10}}\end{picture}}

\newsavebox{\war}\savebox{\war}(30,30){\begin{picture}(30,30)
\thicklines\put(0,15){\line(2,1){20}}\put(0,15){\line(2,-1){20}}
\put(20,5){\line(0,1){20}}\end{picture}}

\newsavebox{\sar}\savebox{\sar}(30,30){\begin{picture}(30,30)
\thicklines\put(15,0){\line(1,2){10}}\put(15,0){\line(-1,2){10}}
\put(5,20){\line(1,0){20}}\end{picture}}

\newsavebox{\ear}\savebox{\ear}(30,30){\begin{picture}(30,30)
\thicklines\put(10,5){\line(0,1){20}}\put(10,5){\line(2,1){20}}
\put(10,25){\line(2,-1){20}}\end{picture}}

\newsavebox{\mybox}\savebox{\mybox}(350,350){
\begin{picture}(350,350)
\thicklines
\multiput(30,112.5)(15,0){16}{\line(1,0){10}}
\multiput(30,187.5)(15,0){16}{\line(1,0){10}}
\multiput(112.5,30)(0,15){16}{\line(0,1){10}}
\multiput(187.5,30)(0,15){16}{\line(0,1){10}}
\multiput(75,150)(75,75){2}{\circle*{10}}
\multiput(150,75)(75,75){2}{\circle*{10}}
\multiput(74,150)(1,0){3}{\line(1,1){75}}
\multiput(74,150)(1,0){3}{\line(1,-1){75}}
\multiput(149,75)(1,0){3}{\line(1,1){75}}
\multiput(149,225)(1,0){3}{\line(1,-1){75}}
\multiput(134,210)(1,0){3}{\line(-2,-1){20}}
\multiput(136,210)(0,1){3}{\line(-1,-2){10}}
\multiput(209,165)(1,0){3}{\line(-2,1){20}}
\multiput(211,165)(0,-1){3}{\line(-1,2){10}}
\multiput(134,90)(1,0){3}{\line(-2,1){20}}
\multiput(136,90)(0,-1){3}{\line(-1,2){10}}
\multiput(166,90)(-1,0){3}{\line(2,1){20}}
\multiput(164,90)(0,-1){3}{\line(1,2){10}}
\end{picture}}

\newsavebox{\stars}\savebox{\stars}(350,350){
\begin{picture}(350,350)
\thicklines
\multiput(40,112.5)(15,0){16}{\line(1,0){10}}
\multiput(40,187.5)(15,0){16}{\line(1,0){10}}
\multiput(187.5,30)(0,15){16}{\line(0,1){10}}
\multiput(112.5,30)(0,15){16}{\line(0,1){10}}
\multiput(75,225)(150,0){2}{\circle*{10}}
\multiput(75,75)(150,0){2}{\circle*{10}}
\put(150,150){\circle*{10}}
\multiput(149,150)(1,0){3}{\line(1,1){75}}
\multiput(149,150)(1,0){3}{\line(1,-1){75}}
\multiput(74,75)(1,0){3}{\line(1,1){75}}
\multiput(74,225)(1,0){3}{\line(1,-1){75}}
\multiput(110,160)(75,-75){2}{\usebox{\sear}}
\put(110,110){\usebox{\near}}\put(160,160){\usebox{\swar}}
\end{picture}}

\newsavebox{\starss}\savebox{\starss}(350,350){
\begin{picture}(350,350)
\thicklines
\multiput(40,112.5)(15,0){16}{\line(1,0){10}}
\multiput(40,187.5)(15,0){16}{\line(1,0){10}}
\multiput(187.5,30)(0,15){16}{\line(0,1){10}}
\multiput(112.5,30)(0,15){16}{\line(0,1){10}}
\multiput(75,225)(150,0){2}{\circle*{10}}
\multiput(75,75)(150,0){2}{\circle*{10}}
\put(150,150){\circle*{10}}
\multiput(149,150)(1,0){3}{\line(1,1){75}}
\multiput(149,150)(1,0){3}{\line(1,-1){75}}
\multiput(74,75)(1,0){3}{\line(1,1){75}}
\multiput(74,225)(1,0){3}{\line(1,-1){75}}
\multiput(110,160)(75,-75){2}{\usebox{\sear}}
\put(185,185){\usebox{\near}}\put(85,85){\usebox{\swar}}
\end{picture}}

\def\b{\beta}
\def\d{\delta}
\def\g{\gamma}
\def\a{\alpha}
\def\s{\sigma}
\def\t{\tau}
\def\l{\lambda}
\def\e{\epsilon}
\def\r{\rho}
\def\d{\delta}
\def\wid{\widehat}
\def\ds{\displaystyle}
\def\be{\begin{equation}}
\def\ee{\end{equation}}
\def\beq{\begin{eqnarray}}
\def\eeq{\end{eqnarray}}
\def\ov{\overline}
\def\om{\omega}
\hfill{\large February, 1999}

\vspace{2cm}
\centerline{\LARGE Bethe ansatz for the three-layer Zamolodchikov model}
\vspace{1cm}

\centerline{\Large H.E. Boos\footnote{
On leave of absence 
from Institute for High Energy Physics, Protvino, 142284, Russia,
E-mail: boos@avzw02.physik.uni-bonn.de}}
\centerline{\large
Physikalisches Institut der Universit{\"a}t Bonn}
\centerline{\large 53115, Bonn,
Germany}

\vspace{1cm}
\centerline{\Large V.V. Mangazeev\footnote{E-mail:
vladimir@maths.anu.edu.au}}
\centerline{\large Centre for Mathematics and its Applications,}
\centerline{\large  School of Mathematical Sciences,}
\centerline{\large The Australian National University,}
\centerline{\large Canberra, ACT 0200, Australia}

\vspace{1cm}
\begin{abstract}
This paper is a continuation of our previous work \cite{BM}.
We obtain two more functional relations for the eigenvalues of
the transfer matrices for the $sl(3)$ chiral Potts model at
$q^2=-1$. This model, up to a modification of boundary conditions,
is equivalent to the three-layer three-dimensional Zamolodchikov
model. From these relations we derive the Bethe ansatz equations.
\end{abstract}

\newpage

\section{Introduction}

One of the open problems in the theory of 
integrable statistical systems is to construct
the Bethe ansatz technique for 
three-dimensional integrable lattice models. 
A construction of such a model is connected
with the problem of solving  tetrahedron
equations \cite{BS1,JM}
which insure integrability of a three-dimensional model.
These  are a system of thousands of 
equations in the simplest nontrivial case. Hence, the problem of
solving them is very difficult. 

There are only a few known integrable three-dimensional models
which are interesting from the physical point of view.
The first nontrivial example of such a model was proposed
by Zamolodchikov in 1980 in  \cite{Z1,Z2}.
The tetrahedron equations for the Zamolodchikov model
were proved by Baxter in \cite{BT}. 

Bazhanov and Stroganov \cite{BS2} observed that 
the Zamolodchikov model and the three-dimensional 
free-fermion model were "weakly equivalent", i.e., 
the free energy of the Zamolodchikov model and
the free-fermion model satisfied the same symmetry and inversion relations.
The assumption that analytical properties of the free energy
were also the same resulted in a coincidence of the free energy
for the Zamolodchikov model and the free-fermion model.
In 1986 Baxter \cite{Bax} calculated the partition function
for the Zamolodchikov model with some modification of 
boundary conditions
for the case of the infinite cubic lattice 
and for the case of the lattice which is 
infinite in two dimensions and finite
in the third one. His result was similar to the result by Bazhanov
and Stroganov for the free-fermion model but not the same.
Namely, the partition function for the Zamolodchikov model was made up
of a sum of two parts. The first part coincided with the partition
function of the free fermion model and
had the usual analytical properties for two-dimensional models. 
The second part was expressed
in terms of the Euler dilogarithm function and had the cut
in the complex plane. Therefore, the assumption that the free energy
for the Zamolodchikov
model and  the free-fermion model had the same analytical properties
was incorrect.
However, the similarity of these results was remarkable.
Later Baxter and Quispel in \cite{BQ}
tried to clarify this fact.
Namely, they constructed the Hamiltonian for the two- and three-layer 
Zamolodchikov model. The two-layer case turned out to correspond to
the two-dimensional free-fermion model. 
The Hamiltonian for the three-layer
case contained cubic interaction terms and seemed not to
be the Hamiltonian for the free-fermion model.

Another important step in the theory of the 
integrable three-dimensional models was done
by Baxter and Bazhanov in 1992. Namely,
they observed \cite{BB} that
the $sl(n)$ chiral Potts model at $q^{2N}=1$ \cite{BKMS,DJMM} 
was equivalent to the $n$-layer three-dimensional model
which turned out to be the $N$-state generalization of the 
Zamolodchikov model. 
It was also mentioned that as for the Zamolodchikov model
this equivalence is valid only
up to some modification of boundary conditions which
should not effect the partition function in the thermodynamic
limit. The partition function for the Baxter-Bazhanov model
was calculated in their next paper \cite{BB1}. The result
appeared to be connected in a remarkably simple way with
that for the Zamolodchikov model.

We hope that a development of the Bethe ansatz technique
for the Zamo-\\
lodchikov and Baxter-Bazhanov models could shed
a new light on the problems discussed above.
Since the $n$-layer case of the Zamolodchikov model
is equivalent to the $sl(n)$ chiral Potts model at $q^2=-1$,
up to a modification of boundary conditions,
we can try to construct a Bethe ansatz for the $sl(n)$ chiral
Potts model.

The Bethe ansatz technique is usually applicable to the study
of effects connected with the finite size of a lattice.
Therefore, there is a good chance that it will  be useful for 
an investigation
of the finite size corrections and the excitations.

Our first step is to develop this program for the
three-layer case of the Zamolodchikov model with modified
boundary conditions, i.e.,
for the $sl(3)$ chiral Potts model at $q^2=-1$.

This work is a continuation of our previous 
paper \cite{BM} where some functional relations for the 
$sl(3)$ chiral Potts model at $q^2=-1$ were 
derived and the nested Bethe ansatz was constructed
in the particular case when the vertical rapidity parameters
coincide. Unfortunately, we did not succeed
in solving these functional relations. 
Our goal here is to derive other functional relations and 
to obtain from them
the Bethe ansatz equations  for the general case.

The paper is organized as follows. In Section 2 we recall 
the basic formulations of the $sl(3)$ chiral Potts model
and its correspondence to the modified three-layer Zamolodchikov
model. In Section 3 we fix the definitions of the transfer matrices
and discuss some of their simple properties. In Section 4
we give two functional
relations for the eigenvalues of the transfer matrices. 
In Section 5 we obtain the Bethe ansatz equations.
In the last section we give a brief discussion of the obtained results
and directions for further investigation.
In the Appendix we outline the basic steps of the proof of one of the
functional relations.

\section{Basic formulations.}
The basic formulation of the Zamolodchikov model and
it's generalization, the Baxter-Bazhanov model can be found
in papers \cite{Z1,Z2} and \cite{BB}. In the last paper
it was observed that the Boltzmann weights for the $sl(n)$ 
chiral Potts model at $q^{2N}=1$ were a product of the $n$ more simple
weights (see formulae 
(~3.7~-~3.13~) of \cite{BB}). 
Hence, the "star" weight for $sl(n)$ chiral Potts
model appeared to be a product of the $n$ weight functions
interpreted as the Boltzmann weights for some
$N$-state three-dimensional model:

\begin{picture}(700,310)
\phantom{a}
\put(-20,40){
\begin{picture}(640,300)
\thicklines
\multiput(100,60)(-1,1){2}{\line(1,1){60}}
\multiput(160,120)(-1,1){2}{\line(1,1){60}}
\multiput(100,180)(-1,-1){2}{\line(1,-1){60}}
\multiput(160,120)(-1,-1){2}{\line(1,-1){60}}
\put(100,60){\circle*{10}}
\put(220,60){\circle*{10}}
\put(100,180){\circle*{10}}
\put(220,180){\circle*{10}}
\put(160,120){\circle*{10}}
\put(20,30){$(\a_1\ldots\a_n)$}
\put(180,30){$(\b_1\ldots\b_n)$}
\put(20,200){$(\d_1\ldots\d_n)$}
\put(180,200){$(\g_1\ldots\g_n)$}
\put(180,110){$(\s_1\ldots\s_n)$}
\put(60,100){$\sum_{\{\s\}}$}
\put(350,120){\large$=$}
\put(-70,-10){
\begin{picture}(600,300)
\put(580,50){\line(1,0){120}}
\multiput(580,50)(-20,20){4}{\line(0,1){120}}
\multiput(580,170)(-20,20){4}{\line(1,0){120}}
\put(700,50){\line(0,1){120}}
\put(580,50){\line(-1,1){20}}
\put(580,170){\line(-1,1){20}}
\put(700,170){\line(-1,1){20}}
\put(540,90){\line(-1,1){20}}
\put(540,210){\line(-1,1){20}}
\put(660,210){\line(-1,1){20}}
\multiput(555,75)(-5,5){3}{\circle*{1}}
\multiput(555,195)(-5,5){3}{\circle*{1}}
\multiput(675,195)(-5,5){3}{\circle*{1}}
\put(560,30){$\a_1$}
\put(540,50){$\a_2$}
\put(505,85){$\a_n$}
\put(485,105){$\a_1$}
\put(705,30){$\b_1$}
\put(590,150){$\d_1$}
\put(495,235){$\d_1$}
\put(710,160){$\g_1$}
\put(690,185){$\g_2$}
\put(655,225){$\g_n$}
\put(630,250){$\g_1$}
\multiput(540,65)(-5,5){3}{\circle*{1}}
\multiput(690,205)(-5,5){3}{\circle*{1}}
\end{picture}}
\put(310,-10){\bf Figure 1.}
\end{picture}}
\end{picture}

Each weight of the product in the RHS depends on the eight spins 
with $N$ possible values.
For the case $N=2$ this model turned out to be just the Zamolodchikov model
rewritten by Baxter in the "interaction-round-cube" form \cite{BT}. 
As it was already mentioned in Introduction this equivalence is valid
up to some modification of boundary conditions. 

Since we study the three-layer case
of the Zamolodchikov model we need to consider the $sl(3)$ chiral Potts
model at $q^2~=~-~1$. The basic notations of this model were adduced 
in papers \cite{BKMS,KMN} or in \cite{BB}. 
It can also be found in our previous paper \cite{BM}
but to be independent here we give 
some necessary basic definitions  below.

The model is formulated on the square lattice (see Figure 2)

\begin{picture}(600,520)
\put(50,50){
\begin{picture}(600,380)
\multiput(0,150)(150,0){3}{\usebox{\stars}}
\multiput(0,0)(150,0){3}{\usebox{\stars}}
\multiput(20,172.5)(0,150){2}{\usebox{\war}}
\multiput(570,97.5)(0,150){2}{\usebox{\ear}}
\multiput(176.5,10)(150,0){3}{\usebox{\sar}}
\multiput(101.5,410)(150,0){3}{\usebox{\nar}}
\end{picture}}
\put(310,20){\bf Figure 2.}
\end{picture}

The interaction is
defined by two types of the weight functions $\overline W_{pq}(\a,\b)$ 
and  $(\overline W_{qp}(\a,\b))^{-1}$ which depend on the neighbouring
spin variables and spectral parameters. The rule how to choose these
weights is shown on Figure~3:

\begin{picture}(700,300)
\put(10,80)
{\begin{picture}(300,150)
\thicklines
\multiput(0,0)(30,30){5}{\line(1,1){20}}
\multiput(150,0)(-30,30){5}{\line(-1,1){20}}
\put(120,120){\usebox{\near}}\put(0,120){\usebox{\nwar}}
\multiput(75,0)(0,150){2}{\circle*{10}}
\multiput(74,0)(1,0){3}{\line(0,1){150}}
\multiput(74,120)(1,0){3}{\line(-1,-2){10}}
\multiput(74,120)(1,0){3}{\line(1,-2){10}}
\put(0,30){\large $p$}\put(150,30){\large $q$}
\put(90,0){\large $\a$}\put(90,150){\large $\b$}
\put(150,75){\bf {$=\quad \overline W_{pq}(\a,\b)$}}
\end{picture}}
\put(280,30){\bf Figure 3.}
\put(400,80)
{\begin{picture}(300,150)
\thicklines
\multiput(0,150)(30,-30){5}{\line(1,-1){20}}
\multiput(150,150)(-30,-30){5}{\line(-1,-1){20}}
\put(0,0){\usebox{\swar}}\put(120,0){\usebox{\sear}}
\multiput(75,0)(0,150){2}{\circle*{10}}
\multiput(74,0)(1,0){3}{\line(0,1){150}}
\multiput(74,120)(1,0){3}{\line(-1,-2){10}}
\multiput(74,120)(1,0){3}{\line(1,-2){10}}
\put(0,30){\large $p$}\put(150,30){\large $q$}
\put(90,0){\large $\a$}\put(90,150){\large $\b$}
\put(150,75){{$=\quad{\displaystyle 1 \over 
{\displaystyle\phantom{I}{ \overline W_{qp}(\a,\b)}^{\phantom{I}}}}$}}
\end{picture}}
\end{picture}

The Boltzmann weights depend on the rapidity
parameters.
Each rapidity variable is represented by three $2$-vectors 
($h_i^+(p)$, $h_i^-(p)$), $i=1,2,3$
which specify the point $p$ of the algebraic curve $\Gamma$ 
defined by relations
\be
\pmatrix{h_i^+(p)^2\cr h_i^-(p)^2\cr}=K_{ij}
\pmatrix{h_j^+(p)^2\cr h_j^-(p)^2\cr},\quad \forall i,j=1,2,3,  \label{l1}
\ee
where $K_{ij}$ are $2\times2$ complex matrices of moduli satisfying
\be
detK_{ij}=1,\quad K_{ii}=K_{ij}K_{jk}K_{ki}=1    \label{l2}                     
\ee
and indices $i,j,k$ take values $1,2,3$ {\it modulo} $3$.

Further we will need the automorphism $\tau$ on the 
curve $\Gamma$ defined as follows
\be
h_j^+(\tau(p))=h_j^+(p),\quad
h_j^-(\tau(p))=- h_j^-(p), \quad j=1,2,3. \label{l2b}
\ee

The curve $\Gamma$ can be defined in a different way which is also useful.
Namely, for two arbitrary points $p$ and $q$ on this curve  the following
combination
\be 
\Delta_{pq}\, = \, {h_i^{+}(p)}^2\, {h_i^{-}(q)}^2
\,-\,{h_i^{-}(p)}^2\, 
{h_i^{+}(q)}^2,
\label{delta}
\ee
should be the same for all $i=1,2,3$.
It is easy to see that both these definitions are equivalent to each other.

The Boltzmann weights depend also on spin variables.
Each spin variable is described by a  two-vector
\be
\a\equiv(\a_1,\a_2),\quad\a_i\in Z_2\quad i=1,2. \label{var}
\ee

Then  the function $\ov W_{pq}(\a,\b)$, $\a,\b\in Z_2\times Z_2$ is 
defined as
\be
\ov W_{p,q}(\a,\b)=(-1)^{Q(\a,\b)}g_{pq}(0,\a-\b), \label{l4}
\ee
where 
\be
Q(\a,\b)=\b_1(\b_1-\a_1)+\b_2(\b_1-\a_1+\b_2-\a_2),
\quad\a,\b\in Z_2\times Z_2\label{l6}
\ee
and the function $g_{pq}(0,\a)$ has the following form
\be
g_{pq}(0,\a)={\ds\prod_{\b=0}^{\a_1+\a_2-1}(h_3^+(p)h_3^-(q)-h_3^+(q)h_3^-(p)
(-1)^{-\b})\over  \ds
\prod_{i=1}^2
\prod_{\b_i=0}^{\a_i-1}(h_i^+(p)h_i^-(q)-h_i^+(q)h_i^-(p)(-1)^{1+\b_i})}.
\label{l7}
\ee

We choose a normalisation of $\ov W_{pq}(\a,\b)$ as 
\be
\ov W_{pq}(0,0)=1. \label{l8}
\ee
Then it is easy to see that
\be
\ov W_{pp}(\a,\b)=\ov\d_{\a,\b} \label{l9}
\ee
where 
\be
    \overline{\delta}_{\a,\b}\equiv\cases{1,&$\a=\b\pmod{2};$\cr
                                           0,& otherwise.\cr}  \label{l10}
\ee
The function $\ov W_{pq}(\a,\b)$ satisfies the inversion relation
\be
\sum_{\b\in Z_2\times Z_2}\overline W_{pq}(\a,\b)\overline W_{qp}(\b,\g)=
                \overline\delta_{\a,\g}\Phi_{pq},                 \label{l11}
\ee
where the inversion factor $\Phi_{pq}$ is given by:
\be
\Phi_{pq}\,=\,{{4\,E_{pq}}\over{D_{pq}}},                     \label{l12}
\ee
and
\be
E_{pq}\,=\,\prod_{i=1}^3 h_i^{+}(p)h_i^{-}(q)\;+\;
\prod_{i=1}^3 h_i^{-}(p)h_i^{+}(q),\label{E}
\ee
\be
D_{pq}\,=\,\prod_{i=1}^3(h_i^{+}(p)h_i^{-}(q)+h_i^{-}(p)h_i^{+}(q)).
\label{D}
\ee

As it was shown in \cite{RMK} (see \cite{BB} for 
details)  the Boltzmann weights 
$\ov W$ satisfy
the "star-star" relation  
which provide the integrability of the $sl(n)$ chiral Potts
model. This relation looks as
\be
{\ov W_{p'p}(\d,\a)\over \ov W_{p'p}(\g,\b)} 
W_{qq'}^{pp'}(\a,\b,\g,\d)=
\wid W_{q'q}^{p'p}(\a,\b,\g,\d)
{\ov W_{q'q}(\a,\b)\over \ov W_{q'q}(\d,\g)},
\label{l23}
\ee
where two ``star'' weights are defined as follows
\be
W_{q'q}^{p'p}(\a,\b,\g,\d)=\sum_\s
{\ov W_{pq}(\a,\s)\ov W_{p'q'}(\g,\s)\ov W_{q'p}(\s,\b)\over
\ov W_{p'q}(\d,\s)},                        \label{l21}
\ee
\be
\wid W_{q'q}^{p'p}(\a,\b,\g,\d)=\sum_\s
{\ov W_{pq}(\s,\g)\ov W_{p'q'}(\s,\a)\ov W_{q'p}(\d,\s)\over
\ov W_{p'q}(\s,\b)}.                        \label{l22}
\ee

The objects defined in (\ref{l21}-\ref{l22}) are
the "star"-weights. 
As it was mentioned above these weights correspond to the three-layer 
case of the Zamolodchikov model. To be exact for the general case of 
the rapidity variables $h_i^{\pm}(p)$ satisfying (\ref{l1}) 
the corresponding Zamolodchikov model is
inhomogeneous in the third direction. 
In fact, we will be interested mainly in the homogeneous case:
\be
h^+_i(p)\,=\,1,\quad\quad h^-_i(p)\,=\,p.
\label{trig}
\ee
It is easy to see that the defining relations (\ref{delta}) are trivially
satisfied. Therefore we do not need to work with the high genus curve
$\Gamma$. 

As it was pointed out in \cite{BB1} the rapidity variables can be 
parameterized
in terms of the spherical angles and excesses $\theta_1,\theta_2,a_3$
\be
{q'\over q}\,=\,-i\,\tan{{\theta_2}\over{2}},\quad
{p\over p'}\,=\,i\,\tan{{\theta_1}\over{2}},\quad
{p\over q}\,=\,e^{-i{a_3\over2}}\sqrt{\tan{{\theta_1}\over2}
\tan{{\theta_2}\over2}}.
\label{spher}
\ee

\section{Transfer matrices}

Here we use slightly different definitions of the
transfer matrices comparing with \cite{BM}:
\be
{T(p;q,q')}_{i_1,\ldots,i_N}^{j_1,\ldots,j_N}\,=\,\prod_{k=1}^N
{
{{\ov W_{pq}}(i_k,j_k)\,{\ov W_{q'p}}(j_k,i_{k+1})}
\over
{{\ov W_{q'q}}(i_k,i_{k+1})},
}
\label{T}
\ee
\be
{\ov T(p;q,q')}_{i_1,\ldots,i_N}^{j_1,\ldots,j_N}\,=\,\prod_{k=1}^N
{
{{\ov W_{q'q}}(j_k,j_{k+1})\,{\ov W_{pq'}}(j_{k+1},i_k)}
\over
{{\ov W_{pq}}(j_k,i_k)}
}
\label{TT}
\ee
which are shown  on Figures 4 and 5:

\begin{picture}(700,250)
\put(-20,0){\begin{picture}(600,250)
\thicklines
\multiput(160,100)(0,1){2}{\line(1,0){160}}
\put(160,100){\vector(1,0){120}}
\multiput(320,100)(0,1){2}{\line(1,0){160}}
\put(320,100){\vector(1,0){120}}
\multiput(560,100)(0,1){2}{\line(1,0){160}}
\put(560,100){\vector(1,0){120}}
\multiput(160,100)(-1,1){2}{\line(1,1){80}}
\put(160,100){\vector(1,1){60}}
\multiput(240,180)(-1,-1){2}{\line(1,-1){80}}
\put(240,180){\vector(1,-1){60}}
\multiput(320,100)(-1,1){2}{\line(1,1){80}}
\put(320,100){\vector(1,1){60}}
\multiput(400,180)(-1,-1){2}{\line(1,-1){80}}
\put(400,180){\vector(1,-1){60}}
\multiput(560,100)(-1,1){2}{\line(1,1){80}}
\put(560,100){\vector(1,1){60}}
\multiput(640,180)(-1,-1){2}{\line(1,-1){80}}
\put(640,180){\vector(1,-1){60}}
\put(0,120){$T(p;q,q')\quad :$}
\put(160,100){\circle*{10}}
\put(320,100){\circle*{10}}
\put(480,100){\circle*{10}}
\put(560,100){\circle*{10}}
\put(720,100){\circle*{10}}
\put(240,180){\circle*{10}}
\put(400,180){\circle*{10}}
\put(640,180){\circle*{10}}
\multiput(500,100)(10,0){5}{\circle{1}}
\multiput(180,140)(40,0){14}{\line(1,0){20}}
\put(700,140){\vector(1,0){20}}
\put(150,135){$p$}
\multiput(170,170)(30,-30){4}{\line(1,-1){20}}
\put(190,150){\vector(-1,1){20}}
\multiput(200,60)(30,30){4}{\line(1,1){20}}
\put(220,80){\vector(-1,-1){20}}

\multiput(330,170)(30,-30){4}{\line(1,-1){20}}
\put(350,150){\vector(-1,1){20}}
\multiput(360,60)(30,30){4}{\line(1,1){20}}
\put(380,80){\vector(-1,-1){20}}

\multiput(570,170)(30,-30){4}{\line(1,-1){20}}
\put(590,150){\vector(-1,1){20}}
\multiput(600,60)(30,30){4}{\line(1,1){20}}
\put(620,80){\vector(-1,-1){20}}

\put(200,40){$q'$}
\put(280,40){$q$}
\put(360,40){$q'$}
\put(440,40){$q$}
\put(600,40){$q'$}
\put(680,40){$q$}

\put(150,70){$i_1$}
\put(310,70){$i_2$}
\put(470,70){$i_3$}
\put(550,70){$i_N$}
\put(710,70){$i_1$}
\multiput(510,75)(10,0){3}{\circle*{1}}
\put(230,200){$j_1$}
\put(390,200){$j_2$}
\put(630,200){$j_N$}
\multiput(490,205)(10,0){7}{\circle*{1}}
\put(350,0){\bf Figure 4.}
\end{picture}}
\end{picture}

\begin{picture}(700,250)
\put(-20,0){\begin{picture}(600,250)
\thicklines
\multiput(160,180)(0,1){2}{\line(1,0){160}}
\put(160,180){\vector(1,0){120}}
\multiput(320,180)(0,1){2}{\line(1,0){160}}
\put(320,180){\vector(1,0){120}}
\multiput(560,180)(0,1){2}{\line(1,0){160}}
\put(560,180){\vector(1,0){120}}

\multiput(160,180)(-1,-1){2}{\line(1,-1){80}}
\put(160,180){\vector(1,-1){60}}
\multiput(240,100)(-1,1){2}{\line(1,1){80}}
\put(320,180){\vector(-1,-1){60}}

\multiput(320,180)(-1,-1){2}{\line(1,-1){80}}
\put(320,180){\vector(1,-1){60}}
\multiput(400,100)(-1,1){2}{\line(1,1){80}}
\put(480,180){\vector(-1,-1){60}}

\multiput(560,180)(-1,-1){2}{\line(1,-1){80}}
\put(560,180){\vector(1,-1){60}}
\multiput(640,100)(-1,1){2}{\line(1,1){80}}
\put(720,180){\vector(-1,-1){60}}

\put(0,120){$\ov T(p;q,q')\quad :$}
\put(160,180){\circle*{10}}
\put(320,180){\circle*{10}}
\put(480,180){\circle*{10}}
\put(560,180){\circle*{10}}
\put(720,180){\circle*{10}}
\put(240,100){\circle*{10}}
\put(400,100){\circle*{10}}
\put(640,100){\circle*{10}}
\multiput(500,180)(10,0){5}{\circle{1}}
\multiput(170,140)(40,0){14}{\line(1,0){20}}
\put(190,140){\vector(-1,0){20}}
\put(150,135){$p$}

\multiput(200,220)(30,-30){4}{\line(1,-1){20}}
\put(260,200){\vector(1,1){20}}
\multiput(170,110)(30,30){4}{\line(1,1){20}}
\put(290,130){\vector(1,-1){20}}

\multiput(360,220)(30,-30){4}{\line(1,-1){20}}
\multiput(330,110)(30,30){4}{\line(1,1){20}}
\put(420,200){\vector(1,1){20}}
\put(450,130){\vector(1,-1){20}}

\multiput(600,220)(30,-30){4}{\line(1,-1){20}}
\multiput(570,110)(30,30){4}{\line(1,1){20}}
\put(660,200){\vector(1,1){20}}
\put(690,130){\vector(1,-1){20}}

\put(170,80){$q$}
\put(290,80){$q'$}

\put(330,80){$q$}
\put(450,80){$q'$}

\put(570,80){$q$}
\put(690,80){$q'$}

\put(230,70){$i_1$}
\put(390,70){$i_2$}
\put(630,70){$i_N$}
\multiput(490,75)(10,0){7}{\circle*{1}}
\put(150,200){$j_1$}
\put(310,200){$j_2$}
\put(470,200){$j_3$}
\put(550,200){$j_N$}
\put(710,200){$j_1$}
\multiput(510,205)(10,0){3}{\circle*{1}}
\put(350,20){\bf Figure 5.}
\end{picture}}
\end{picture}

\noindent
where we imply the cyclic boundary conditions $i_{N+1}=i_1$
and $j_{N+1}=j_1$.

We note that these definitions differ from the previous ones
just by the diagonal equivalence transformation. Of course, it does not
effect the partition function.

Below we shall use more simple notations $T_p = T(p;q,q')$
and ${\ov T_p} ={\ov T(p;q,q')}$ assuming that 
the rapidities $q$ and $q'$ are fixed.
Due to  (\ref{l23}) these transfer matrices 
$T_p$ and $\ov T_p$ commute.
Namely, for two arbitrary rapidities $p$ and $p'$
\be
[T_p,T_{p'}]\,=\, [{\ov T_p},{\ov T_{p'}}]\,=\,
[T_p,{\ov T_{p'}}]\,=\,0.
\label{com}
\ee

One can consider some limiting cases. 
From 
(\ref{l4}-\ref{l8}) we can conclude
that if $q'\rightarrow p$ we have
\be
T_p\,=\,X^{-1},\quad {\ov T}_p\,=\,X,
\label{lim1}
\ee
where $X$ is the shift-operator:
\be
X_{i_1\ldots\i_N}^{j_1\ldots\j_N}\,=\,\prod_{k=1}^N\d_{i_k,j_{k+1}};
\label{shift}
\ee
if $q\rightarrow p$ then 
\be
T_p\,=\,I,
\label{lim2}
\ee
while $\ov T$ has the singular matrix elements.

\section{Functional relations}

Further we will consider only the case of
the homogeneous  three-layer Zamo-\\
lodchikov model. 
Due to the commutation relations (\ref{com}) we can diagonalize 
the transfer matrices $T_p$ and $\ov T_p$ simultaneously.

Let us denote eigenvalues of $T_p$ and $\ov T_p$ by
$t(p)$ and $\ov t(p)$ where we omit a dependence on $q$ and $q'$.

In Appendix we outline the proof of the following couple of functional
relations 
$$
{\ov t(p)}\;\;t(p)\;\;{\ov t(-p)}\;\;t(-\om p)\,=\,
$$
$$
\phi_0^N\;{\ov t(p)}\,t(-\om p)\,+\,
\phi_1^N\;{\ov t(-p)}\,t(-\om p)\,+\,
\phi_2^N\;{\ov t(p)}\,t(\om p)\,+\,
\phi_3^N\;{\ov t(-p)}\,t(\om p)
$$
\be
\quad
\label{fe}
\ee
and 
$$
{\ov t(-\om p)}\;\;t(-p)\;\;{\ov t(p)}\;\;t(p)\,=\,
$$
$$
{\phi'}_0^N\;{\ov t(-\om p)}\,t( p)\,+\,
{\phi'}_1^N\;{\ov t(-\om p)}\,t(-p)\,+\,
{\phi'}_2^N\;{\ov t(\om p)}\,t(p)\,+\,
{\phi'}_3^N\;{\ov t(\om p)}\,t(-p),
$$
\be
\quad
\label{fe1}
\ee
where
\be
\phi_0\,=\,4\,{{(p\,+\,\om q)\;(p\,+\,\om^{-1}q)}\over{(p\,+\,q)^2}},\quad
\phi_1\,=\,4\,{{(p\,+\,\om q')\;(p\,+\,\om^{-1} q')}\over{(p\,+\,q')^2}},
\ee
\beq
\phi_2\,=\,4\,
{{(p\,-\,q)\;(p\,+\,\om^{-1} q)^2\;(p\,+\,\om^{-1} q')\;(p\,-\,\om q')}
\over
{(p\,-\,\om^2 q)\;(p\,+\,q)^2\;(p\,+\,q')\;(p\,-\,\om^{-1} q')}},&&
\nonumber\\
\phi_3\,=\,4\,
{{(p\,-\,q')\;(p\,+\,\om^{-1} q')^2\;(p\,+\,\om^{-1} q)\;(p\,-\,\om q)}
\over
{(p\,-\,\om^2 q')\;(p\,+\,q')^2\;(p\,+\,q)\;(p\,-\,\om^{-1} q)}}&&
\label{phi}
\eeq
and $\phi'_i$ can be  obtained from $\phi_i$ by the substitution
$q\rightarrow -q$.
Here $\om$ is the root of unity of power three
$$
\om = e^{{{2\pi i}\over{3}}}.
$$
From the limiting cases (\ref{lim1}) and (\ref{lim2}) we have some
initial data:
\be
t(p;q,p)\,=\,\Omega,\quad {\ov t}(p;q,p)\,=\,\Omega^{-1},\quad t(p;p,q')=1
\label{tlim1}
\ee
where $\Omega$ is some root of unity of power $N$:
\be
\Omega^N=1
\label{Omega}.
\ee

From (\ref{fe}) and
(\ref{fe1}) one can see that the pair of functions $t'$ and $\ov t'$:
\be
t'(p;q,q') = {\ov t(p;-q,q')},\quad{\ov t'(p;q,q')} = t(p;-q,q')
\label{t-tt}
\ee
satisfy the same  relations (\ref{fe}) and (\ref{fe1}).
However, it is not true that ${\ov t(p;q,q')}=t(p;-q,q')$ for
all eigenvalues. The transformation (\ref{t-tt}) interchanges 
also the eigenvectors
of the transfer matrices which belong to the same symmetry sector.

The analysis of the eigenvalues $t(p)$ and $\ov t(p)$ shows that
it is convenient to extract some "kinematic" multipliers:
\be
t(p)\,=\,{{2^N}\over{(p\,+\,q)^N\;(p\,+\,q')^N}}\,s(p),
\quad
{\ov t(p)}\,=\,{{2^N}\over{(p\,-\,q)^N\;(p\,+\,q')^N}}\,
{\ov s(p)}
\label{tt}
\ee
where $s(p)$ and $\ov s(p)$ are the polynomials of the degree $n$ in
the variable $p$. 
So far, we have no proof that the degrees of $s(p)$ and $\ov s(p)$ should
be the same. Therefore, we accept it as a conjecture.

Substituting the definitions (\ref{tt}) into (\ref{fe}-\ref{fe1})
we obtain the functional relations for $s(p)$ and $\ov s(p)$:

\beq
&{\ov s(p)}\;\;s(p)\;\;{\ov s(-p)}\;\;s(-\om p)\,=\,&\nonumber\\
&{\l}_0^N\;{\ov s(p)}\,s(-\om p)\,+\,
{\l}_1^N\;{\ov s(-p)}\,s(-\om p)\,+\,
{\l}_2^N\;{\ov s(p)}\,s(\om p)\,+\,
{\l}_3^N\;{\ov s(-p)}\,s(\om p)&\nonumber\\
\label{fes}
\eeq
and
\beq
&{\ov s(-\om p)}\;\;s(-p)\;\;{\ov s(p)}\;\;s(p)\,=\,&\nonumber\\
&{\l'}_0^N\;{\ov s(-\om p)}\,s(p)\,+\,
{\l'}_1^N\;{\ov s(-\om p)}\,s(-p)\,+\,
{\l'}_2^N\;{\ov s(\om p)}\,s(p)\,+\,
{\l'}_3^N\;{\ov s(\om p)}\,s(-p),&\nonumber\\
\label{fes1}
\eeq
where
\beq
&&{\l}_0\;=\;(p\,+\om\,q)\;(p\,+\om^{-1}\,q)\;(p\,+\,q')\;(p\,-\,q'),
\nonumber\\
&&{\l}_1\;=\;(p\,+\om\,q')\;(p\,+\om^{-1}\,q')\;(p\,+\,q)\;(p\,-\,q),
\nonumber\\
&&{\l}_2\;=\;(p\,-\,q)\;(p\,+\om^{-1}\,q)\;(p\,-\,\om q')\;(p\,-\,q'),
\nonumber\\
&&{\l}_3\;=\;(p\,-\,q')\;(p\,+\om^{-1}\,q')\;(p\,-\,\om q)\;(p\,-\,q)
\label{l}
\eeq
and $\l'_i$ can be obtained from $\l_i$ by the substitution
$q\rightarrow -q$.

\section{Bethe ansatz equations}

To construct the Bethe ansatz we consider zeros of the
polynomials $s(p)$ and $\ov s(p)$:
\be
s(p)\,=\,a_n(q,q')\,\prod_{i=1}^n\,(p\,-\,p_i),
\quad
{\ov s(p)}\,=\,\ov a_n(q,q')\,\prod_{i=1}^n\,(p\,-\,{\ov p_i})
\label{ts}
\ee
where the power $n$ takes only two possible values $2 N$ and  $2N-1$.
The functions $a_n$ and $\ov a_n$ should be compatible with the 
initial conditions (\ref{tlim1}). Unfortunately,
it is not easy to 
calculate them explicitly but their product looks very simple:
\be
a_{2N}(q,q')\,\ov a_{2N}(q,q')=4,\quad
a_{2N-1}(q,q')\,\ov a_{2N-1}(q,q')=N\,(q'^2-q^2).
\ee

Now we can set $p$ to be some zero of the LHS of 
(\ref{fes}) and consider the equations which follow from the RHS.
In fact, we have four possibilities to do this:
\be
 p\rightarrow{\ov p_i},\quad 
 p\rightarrow{-\ov p_i},\quad 
 p\rightarrow{-\om^{-1} p_i},\quad
 p\rightarrow{p_i}.
\label{put}
\ee
It is not  difficult to obtain that first three
possibilities give us two different sets of Bethe ansatz equations:
\be
{{f(p_i,\om^{\pm1},-q)}^N\over {f(p_i,\om^{\pm1},-q')}^N}\,=\,
(-1)^{n-1}
\prod_{j=1}^n{p_i+\om^{\mp1}\ov p_j\over p_i-\om^{\mp1}\ov p_j}
\label{bethe1}
\ee
and
\be
{{f(\ov p_i,\om^{\pm1},\;q)}^N\over {f(\ov p_i,\om^{\pm1},-q')}^N}=
(-1)^{n-1}
\prod_{j=1}^n{\ov p_i+\om^{\mp1}p_j\over \ov p_i-\om^{\mp1}p_j}
\label{bethe2}
\ee
where
\be
f(p,x,q)\,=\,{{p\,-\,x\,q}\over{p\,+\,q}}.
\label{f}
\ee
The fourth possibility in (\ref{put}) gives some complicated
compatibility conditions for the solution to the Bethe ansatz
equations (\ref{bethe1}-\ref{bethe2}).
Of course, $p_i$ and $\ov p_i$ are the functions of $q$ and $q'$.
A similar consideration of the second functional relation
leads to the same Bethe ansatz equations (\ref{bethe1}) and 
(\ref{bethe2}).
It is obvious that $s(p)$ and $\ov s(p)$ are homogeneous in $p,q,q'$.
So let
\be 
q=1, \quad p=ix, \quad q'=iy
\ee
where $x,y$ are {\bf real}.

{\bf Conjecture}
\be
\ov s(x,y) = s^*(x,y) \label{r5}
\ee
We checked it numerically for $N=2,3$. Let us set 
\be
p_i = ir_i(y),\quad \ov p_i=i\ov r_i(y)
\ee
Then
\be
s(p) =a(y)i^n\prod_{i=1}^n(x-r_i(y)),\quad
\ov s(p) =\ov a(y)i^n\prod_{i=1}^n(x-\ov r_i(y))
\ee
From (\ref{bethe1}) we obtain
\be \ov a(y) =(-1)^na^*(y), \quad \ov r_i(y)=r_i^*(y),\quad
i=1,\ldots,n
\ee
It is easy to see that (\ref{bethe2}) can be obtained from (\ref{bethe1})
by a complex conjugation. Somehow  this is the ``proof'' of 
the conjecture
 (\ref{r5}). Then we have
\be
\biggl[{(r_i-y)(ir_i+\om^\e)\over (r_i+\om^\e y)(ir_i-1)}\biggr]^N=
(-1)^{n-1}\prod_{j=1}^n{r_i+\om^{-\e}r_j^*\over r_i-\om^{-\e}r_j^*},
\quad \e=\pm1. \label{betr}
\ee
One can obtain  from (\ref{betr}) the set of equations on
absolute values and phases of $r_i$.

\section{Discussion}

In this paper we have only presented the Bethe ansatz equations.
We shall give the detailed analysis of these equations elsewhere.
The technique we use here is in the spirit of the Baxter $Q$-matrix
method \cite{BaxB}. The role of the $Q$-matrices is played by the one-layer
transfer matrices. It corresponds to the result obtained by
Bazhanov and Stroganov in \cite{BS} for the chiral Potts model.
We think that the algebraic Bethe ansatz technique can also be 
developed. However, there are some problems like an appropriate choice
of the reference state which are out of our understanding so far.

We should note that the functional relations we have derived here and
those which were obtained in \cite{BM} can be considered together.
Perhaps the combining of all these relations could give more information
about the eigenvalues $t(p)$ and $\ov t(p)$.

We hope that the result obtained by Baxter 
in \cite{Bax} for the partition function 
of the Zamolodchikov model
on the lattice $\infty\times\infty\times 3$ can be reproduced in
the thermodynamic limit of the Bethe ansatz equations 
(\ref{bethe1}-\ref{bethe2}). We also hope that a standard program of a
study of the excitations and finite size corrections \footnote{
See for example the book
\cite{KBI} and  references therein.}
can be performed. 

We think that the technique described in  Appendix
can be generalized to the $sl(n)$ case. 
In principle, a general procedure seems to be more or less clear. 
However, the technical difficulties one could face can be, of course, 
much more serious.

\section{Acknowledgments}
The authors would like to thank Murray~Batchelor, Vladimir~Bazhanov, 
Rainald~Flume, Vladislav Fridkin,
G{\"u}nter~von~Gehlen, J.M.~Maillet and Vladimir~Rittenberg 
for stimulating discussions and suggestions. HEB would also
like to thank R.~Flume for his kind hospitality in the Physical
Institute of Bonn University.
This research (VVM) has been  supported by 
the Australian Research Council
and (HEB) by the Alexander von Humboldt Foundation.

\appendix

\section*{Appendix}

Here we outline the derivation of  (\ref{fe}). The second relation
(\ref{fe1}) can be obtained in a similar way.
In fact, we derive it for a general case of the
inhomogeneous Zamolodchikov model.
Our key relation for the transfer matrices $T_p$ and $\ov T_p$ we
would like to obtain looks as follows:
\be
{\ov T_p}\,T_p\,{\ov T_{\t(p)}}\,T_{p^{\star}}\,=\,
\Phi_0^N\,{\ov T_p}\,T_{p^{\star}}\,+\,
\Phi_1^N\,{\ov T_{\t(p)}}\,T_{p^{\star}}\,+\,
\Phi_2^N\,{\ov T_p}\,T_{\t(p^{\star})}\,+\,
\Phi_3^N\,{\ov T_{\t(p)}}\,T_{\t(p^{\star})}
\quad
\label{FE}
\ee
where $p^{\star}$ is one of two nontrivial
solutions of the equation
\be
{{H^+_p}\over{H^-_p}}\,=\,-{{H^+_{p^{\star}}}\over{H^-_{p^{\star}}}}
\ee
where
\be
H^{\pm}_p\,=\,\prod_{i=1}^3\,h^{\pm}_i(p),\quad
u_{pq}\,=\,{{\Delta_{pq}}\over{D_{pq}}}
{{D_{\t(p^{\star})q}}\over{\Delta_{\t(p^{\star})q}}},
\quad\quad
v_{pq}\,=\,{{E_{\t(p)q}}\over{\Delta_{\t(p)q}}}
{{\Delta_{p^{\star}q}}\over{D_{p^{\star}q}}}
\label{uv}
\ee
and
\be
\Phi_0\,=\,4\,{{E_{pq}}\over{D_{pq}}},\quad
\Phi_1\,=\,4\,{{E_{pq'}}\over{D_{pq'}}},\quad
\Phi_2\,=\,4\,u_{pq}\,v_{pq'},\quad
\Phi_3\,=\,4\,v_{pq}\,u_{pq'}.
\label{Phi1}
\ee
The function $E, D$ and $\Delta$ are given by the
formulae (\ref{E},\ref{D}) and (\ref{delta}) respectively.

In fact, in our previous paper \cite{BM} we made the first
step. Namely, we expressed the matrix product 
$T_p\,{\ov T_{\t(p)}}$ as a sum of two terms
\footnote{We considered two cases $\l=0,1$
corresponding to two automorhisms $\t^{\l}$
(see formula (3.2) of \cite{BM}). Here we consider only
the case $\l=1$. A consideration of the case $\l=0$ gives
the same result.}.
The first one corresponds to the first term in the RHS of (\ref{FE}).
The second term was written in terms of some $L$-operators.
When the vertical rapidities $q$ and $q'$ coincide this $L$-operator
corresponds to the second fundamental representation $\ov 3$ of
the quantum $sl(3)$ algebra. Therefore  we shall denote it as
$\ov L$. 

Now we have to do the next step. Namely, we should consider the matrix
product:
\be
({\ov T_p} {\ov L_p})_{\{\g\}}^{\{\a\}} \;=\; 
Tr \, \prod_{i=1}^N B^{\a_i}_{\g_i,\g_{i+1}}(q',q;p), 
\label{TL}
\ee
where 
\be
{[B^{\a}_{\g,\d}(q',q;p)]}_{i,j}\;=
\;\sum_{\b}{{{\ov W_{pq'}}(\b,\g)}\over{{\ov W_{pq}}(\b,\d)}}
\,{\ov L_{ij}}(\b,\a),
\label{B}
\ee
$\ov L$ is given by
\be
{\ov L_{i,j}}(\b,\a)\;=\;\sum_{n,m} C(i,n)\,{\ov W_{\t(p)q'}}(\a,n)\,
{\ov W_{q'p}}(n,\b){{{\ov W_{pq}}(\b,m)}\over{{\ov W_{\t(p)q}}(\a,m)}}\,C(j,m),
\label{L}
\ee
and all indices $\a,\b,\g,\d,n,m$ are two-component vectors taking
one of four possible  states $(0,0),(0,1),(1,0),(1,1)$,  $i,j=1,2,3$,

\be
C(2\, k_1+k_2,n)\;=\;{1\over 2} (-1)^{k_1\,n_1\,+\,k_2\,n_2}.
\label{C}
\ee
 
Inserting the identity matrices $3\times 3$ between
each pair of $B$ in the RHS of (\ref{TL})

\be
I\,=\, \sum_{i=1}^3 \phi_R(i,\a)\times \phi_L(i,\a),
\label{id1}
\ee
where 
\beq
&&\phi_L(1,\a)=(1,0,0),\nonumber\\
&&\phi_L(2,\a)=(-(-1)^{\a_1+\a_2},1,0),\nonumber\\
&&\phi_L(3,\a)=(-(-1)^{\a_1},0,1),
\label{phiL1}
\eeq
\beq
&&\phi_R(1,\a)=(1,(-1)^{\a_1+\a_2},(-1)^{\a_1}),\nonumber\\
&&\phi_R(2,\a)=(0,1,0),\nonumber\\
&&\phi_R(3,\a)=(0,0,1),
\label{phiR}
\eeq
one can check that  the transformed matrices: 
\be
{[{\tilde B^{\a}_{\g,\d}}(q',q;p)]}_{ij}
\,=\,\phi_L(i,\g)\,     B^{\a}_{\g,\d}(q',q;p)\,
\phi_R(j,\d)
\label{BB}
\ee
satisfy the following property
\be
{[{\tilde B^{\a}_{\g,\d}}(q',q;p)]}_{21}\;=\;
{[{\tilde B^{\a}_{\g,\d}}(q',q;p)]}_{31}\,=\,0
\label{dec}
\ee
for all possible values of indices
$\a,\g,\d$.

Therefore, we have a decomposition $1+2$.
It is not difficult to check that 
\be
{[{\tilde B^{\a}_{\g,\d}}(q',q;p)]}_{11}\;=\;
\Phi_{pq'}(-1)^{\g_2+\d_2}{{{\ov W_{\t(p)q'}}(\a,\g)}\over
{{\ov W_{\t(p)q}}(\a,\d)}},
\label{BB1}
\ee
where $\Phi_{pq}$ is defined in (\ref{l12}).
In the RHS of formula (\ref{BB1}) we can recognize
the "building block" of the transfer matrix $\ov T_{\t(p)}$.
So, after taking the product and trace as in the RHS of (\ref{TL})
we obtain the second term in (\ref{FE}).

Now let us define $2\times 2$ matrices  with elements
\be
{[{\wid B^{\a}_{\g,\d}}(q',q;p)]}_{ij}
\,=\,{[{\tilde B^{\a}_{\g,\d}}(q',q;p)]}_{i+1,j+1},\quad i,j=1,2.
\label{BB2}
\ee

The matrices ${\wid B^{\a}_{\g,\d}}(q',q;p)$ have the form
of the following matrix product:
\be
{\wid B^{\a}_{\g,\d}}(q',q;p)\,=\,V_{pq'}(\a,\g)\,U_{pq}(\a,\d)
\label{prod}
\ee
where
\be
{[U_{pq}(\a,\d)]}_{ij}\,=\,\sum_{n,m}\chi_L(i;m,\a)
{{{\ov W_{pq}}(m,n)}\over{{\ov W_{pq}}(m,\d)
{\ov W_{\t(p)q}}(\a,n)}}\chi_R(j;n,\d)
\label{U}
\ee
and
\be
{[V_{pq}(\a,\d)]}_{ij}\,=\,\sum_{n,m}\chi_L(i;m,\d)
{\ov W_{\t(p)q}}(\a,m){\ov W_{qp}}(m,n){\ov W_{pq}}(n,\d)\chi_R(j;n,\a).
\label{V}
\ee
Here we use the following notations
\be
\chi_{L(R)}(i;m,\a)\,=\,\sum_{k=1}^3
C(k,m)\,{[\phi_{L(R)}(i+1;\a)]}_k,
\quad i=1,2.
\label{hi}
\ee

It is interesting to note that $U_{pq}$ and $V_{pq}$ satisfy the
property which is similar to that for $\tilde B$ given by (\ref{dec})

\be
{[U_{pq}(\a,\d)]}_{i0}\,=\,{[V_{pq}(\a,\d)]}_{i0}\,=\,0.
\label{dec0}
\ee

In addition, we have
\be
{[U_{pq}(\a,\d)]}_{00}\,=\,-\
{
{\Phi_{pq}}
\over
{4}
}\,(-1)^{\a_2+\d_2}\,{\ov W_{q\t(p)}}(\d,\a)
\label{U00}
\ee
\be
{[V_{pq}(\a,\d)]}_{00}\,=\,-\
{
{\Phi_{pq}}
\over{4}}\,(-1)^{\a_2+\d_2}\,{\ov W_{\t(p)q}}(\a,\d)
\label{V00}
\ee
and $\Phi_{pq}$ is given by (\ref{l12}).


Using the definitions (\ref{U}) and (\ref{V}) we obtain
\be
U_{pq}(\a,\d)\,=(-1)^{\a_1+\a_2}\eta_{pq}(\a_1,\a_2;\d_1,\d_2)
\left(\begin{array}{cc}
-{1\over{\g_2(p,q)}}&{\;\;\;z_{12}(p,q;\a,\d)}\\
-z_{32}(p,q;\a,\d)&\;\;\;1
\label{UU}
\end{array}\right),
\ee
\be
V_{pq}(\a,\d)\,=(-1)^{\a_1+\d_2}\eta_{pq}(\a_1,\a_2;\d_1,\d_2)
\left(
\begin{array}{cc}
{1}&{\;\;\;-z_{12}(p,q;\a,\d)}\\
z_{32}(p,q;\a,\d)&\;\;\;-{1\over{\g_2(p,q)}}
\label{VV}
\end{array}\right),
\ee
where
\be
\g_i\,(p,q)=\,-{{h_i^{+}(p)h_i^{-}(q)}\over{h_i^{-}(p)h_i^{+}(q)}},
\ee
\be
\eta_{pq}(\a_1,\a_2;\d_1,\d_2)=
-{{2\,\Delta_{pq}}\over{D_{pq}}}
\,{h_2^{+}(p)h_2^{-}(q)\over
{\ov W_{pq}}(\a_1+1,\a_2;\d_1,\d_2)}, \label{eta}
\ee
\be
z_{12}(p,q;\a,\d)\, = \, (-1)^{\a_2}
{{\g_1(p,q)\g_2(p,q)\,-\,(-1)^{\a_1+\d_1+\a_2+\d_2}}\over
{(\g_1(p,q)\,-\,(-1)^{\a_1+\d_1})\,\g_2(p,q)}},
\label{z12}
\ee
\be
z_{32}(p,q;\a,\d)\, = \, (-1)^{\d_2}
{{\g_3(p,q)\g_2(p,q)\,-\,(-1)^{\a_1+\d_1}}
\over
{(\g_3(p,q)\,-\,(-1)^{\a_1+\d_1+\a_2+\d_2})\,\g_2(p,q)}}.
\label{z02}
\ee

It is easy to see from (\ref{UU}) and (\ref{VV})
that the matrix $U_{pq}$ is connected with $V_{pq}$ by the 
matrix inversion up to some coefficient:

\beq
&V_{pq}(\a,\b)U_{pq}(\a,\b)=(\g_1(p,q)\g_2(p,q)\g_3(p,q)-1)\times&\nonumber\\
&{
(\g_2(p,q)-(-1)^{\a_2+\d_2})\eta_{p,q}(\a_1,\a_2;\d_1,\d_2)^2
\over
(\g_1(p,q)-(-1)^{\a_1+\d_1})\,(\g_3(p,q)-(-1)^{\a_1+\d_1+\a_2+\d_2})\,
\g_2(p,q)^2}&
\label{inv}
\eeq

The important fact  is a degeneration of 
these matrices which occurs when
\be
\g_1(p,q)\,\g_2(p,q)\,\g_3(p,q)\,=\,1.
\label{degen}
\ee

Using the "star-star" relation 
for the Boltzmann weights $\ov W$ and property (\ref{dec0})
we can prove that the matrices $U$ and $V$ should satisfy the
following important relation:

$$
\sum_{\a}\quad V_{pq'}(\a,\g)\,U_{pq}(\a,\d)\,V_{pp'}(\d',\d)\quad
{\ov W_{q'p'}}(\g',\a){\ov W_{p'q}}(\a,\d')\;\;=
$$

\be
{{{\ov W_{q'q}}(\g',\d')}\over{{\ov W_{q'q}}(\g,\d)}}\quad
\sum_{\b}\quad{\ov W_{p'q}}(\g,\b){\ov W_{q'p'}}(\b,\d)\quad
V_{pp'}(\g',\g)\,U_{pq}(\g',\b)\,V_{pq'}(\d',\b).
\label{urav}
\ee

From this relation we can deduce that choosing the rapidity variable $p'$
in such a way that all matrices $V_{pp'}$ are degenerate we get
the decomposition of the matrices with the following
elements:

\be
{[D_{\g,\d}^{\g',\d'}(q',q;p,p')]}_{ij}\,=\,
\sum_{\a}{[V_{pq'}(\a,\g)\,U_{pq}(\a,\d)]}_{ij}
\quad
{\ov W_{q'p'}}(\g',\a){\ov W_{p'q}}(\a,\d').
\label{DD}
\ee

 It means that we can reduce 
the matrices $D_{\g,\d}^{\g',\d'}(q',q;p,p')$ by the 
quasi-equivalence transformation
to the upper-triangular form. This technique is rather similar to that
which was used by Baxter for a derivation of the Q-matrix equation
for the 6-vertex and 8-vertex models \cite{BaxB}.

So, first we should choose the point on the curve
$p'$ to provide the degeneration of matrices $V_{p,p'}$. 
Therefore, we should fulfil the condition (\ref{degen}) for the pair 
$(p,p')$:

\be
\g_1(p,p')\,\g_2(p,p')\,\g_3(p,p')\,=\,1.
\label{degen1}
\ee

This equation has three solutions (up to some choice of signs). 
One of them 
\be
\Delta_{pp'}=0
\label{delta0}
\ee
corresponds to the automorphism $\t$:
\be
p'=\t(p).
\ee
Two another solutions can be obtained by taking the second power
of (\ref{degen1}) and using (\ref{l1}). In this way we arrive
to the quadratic equation for the coordinates of $p'$
with coefficients depending on coordinates of $p$. Let us
denote its roots as
\be
p_{\pm}\,=\,\t_{\pm}(p).
\label{autpm}
\ee
Let us choose one of these solutions, for example,

\be
p^{\star}\,=\,p_{+}
\label{pstar}
\ee
and set the point $p'$ in the formulae above to be $p^{\star}$.

It is easy to conclude from (\ref{VV}) that up to some 
factor the matrices $V_{pp^{\star}}$ are proportional to

\be
V_{pp^{\star}}(\a,\d)\, \sim \,
\left(\begin{array}{cc}
1&{\;\;\;-(-1)^{\a_2}
{
{\g^{\star}_1\g^{\star}_2-(-1)^{\a_1+\d_1+\a_2+\d_2}}
\over
{(\g^{\star}_1-(-1)^{\a_1+\d_1})\,\g^{\star}_2}
}
}\\
{
{
{(-1)^{\a_2}(\g^{\star}_1-(-1)^{\a_1+\d_1})}
\over
{\g^{\star}_1\g^{\star}_2-(-1)^{\a_1+\d_1+\a_2+\d_2}}
}
}
&\;\;\;-{1\over{\g^{\star}_2}}
\end{array}\right)
\label{VVV}
\ee
where

\be
\g^{\star}_i\,=\,\g_i(p,p^{\star}).
\label{gamstar}
\ee

So, the vectors which provide the decomposition $1+1$ can be chosen
as:

\be
\zeta_L(1;\a,\d)=(1,0),\quad 
\zeta_L(2;\a,\d)=
(-(-1)^{\a_2}
{
{\g^{\star}_1\g^{\star}_2-(-1)^{\a_1+\d_1+\a_2+\d_2}}
\over
{\g^{\star}_1-(-1)^{\a_1+\d_1}}
}
,1),
\label{etaL}
\ee

\be
\zeta_R(2;\a,\d)=(0,1),\quad \zeta_R(1;\a,\d)=(1,
(-1)^{\a_2}
{
{\g^{\star}_1-(-1)^{\a_1+\d_1}}
\over
{\g^{\star}_1\g^{\star}_2-(-1)^{\a_1+\d_1+\a_2+\d_2}}
}
),
\label{etaR}
\ee
which satisfy the natural condition

\be
\sum_{i=1}^2\,\zeta_R(i;\a,\d)\times\zeta_L(i;\a,\d)\,=\,I.
\label{norm}
\ee

Now let us consider the transformed matrices $\wid D_{\g,\d}^{\g',\d'}$:

\be
{[\wid D_{\g,\d}^{\g',\d'}(q',q;p)]}_{ij}\,=\,
\zeta_L(i;\g',\g)\,     D_{\g,\d}^{\g',\d'}(q',q;p,p^{\star})\,
\zeta_R(j;\d',\d)
\label{Dtrans}
\ee
where $i,j=1,2$.

One can check the decomposition property:
\be
{[\wid D_{\g,\d}^{\g',\d'}(q',q;p)]}_{21}\,=\,0.
\label{dec1}
\ee

Now we should study diagonal elements of these matrices
${[\wid D_{\g,\d}^{\g',\d'}(q',q;p)]}_{ii}$, $i=1,2$.

\vspace{0.2cm}

It can be checked that the following expressions for $D_{ii}$ are
valid
\be
{[\wid D_{\g,\d}^{\g',\d'}(q',q;p)]}_{ii}\,=\,\Lambda_i(q',q;p)
{{A_i(\d',\d)}\over{A_i(\g',\g)}}\,
\wid W^{p_i\tau(p^\star)}_{q'q}(\g,\d,\g',\d'),
\label{star}
\ee
where $\wid W^{p_i\tau(p^\star)}_{q'q}(\g,\d,\g',\d')$ is given by
(\ref{l22}) and 
\be
p_1\, = \,p,\quad p_2\,=\,\t(p).
\label{p12}
\ee
For the scalar functions $\Lambda_i$ we have
\be
\Lambda_1(q',q;p)\,=\,4\,v_{q',p}\,u_{q,p},\quad
\Lambda_2(q',q;p)\,=\,4\,u_{q',p}\,v_{q,p}
\label{lam2}
\ee
and the functions $u$ and $v$ are given by (\ref{uv}).
The gauge matrices $A_i$ are given by:
\be
A_1\,=\,
\left(\begin{array}{cccc}
1&{\;\;\;a_1}&{\;\;\;a_1}&{\;\;\;1}\\
{\;\;\;a_1}&{\;\;\;1}&{\;\;\;1}&{\;\;\;a_1}\\
{-a_1}&{-1}&{-1}&{-a_1}\\
{-1}&{-a_1}&{-a_1}&{-1}
\end{array}\right),
\ee
\be
A_2\,=\,
\left(\begin{array}{cccc}
1&{\;\;\;a_2}&{-a_2}&{-1}\\
{-a_2}&{-1}&{\;\;\;1}&{\;\;\;a_2}\\
{\;\;\;a_2}&{\;\;\;1}&{-1}&{-a_2}\\
{-1}&{-a_2}&{\;\;\;a_2}&{1}
\label{A}
\end{array}\right),
\ee
where
\be
a_1\,=\,{{h_3^-(p)\,h_3^+(p^{\star})\,+\,h_3^+(p)\,h_3^-(p^{\star})}
\over{h_3^-(p)\,h_3(p^{\star})\,-\,h_3^+(p)\,h_3^-(p^{\star})}},
\quad a_2\,=\,-1/a_1.
\ee

So, we have succeeded in reducing the four-index objects
${[\wid D_{\g,\d}^{\g',\d'}(q',q;p)]}_{ii}$ to the original
"star"-form.
It is not difficult to observe that after taking the product and trace
we obtain the last two terms in (\ref{FE}).
Using commutation relations (\ref{com}) we can simultaneously
diagonalize the transfer matrices and get the functional relation for the 
eigenvalues. 
Taking into account that for the homogeneous case of the Zamolodchikov 
model the automorphism $\t$ acts just as negating of $p$:
\be
\t(p)\,=\,-p.
\label{tau}
\ee
and rapidity $p^{\star}$ can be taken as
\be
p^{\star}\,=\,-\om\,p
\label{starp}
\ee
we come to (\ref{fe}).

\end{document}